\documentclass[aps, prb, preprint, superscriptaddress]{revtex4}
\usepackage{graphicx}
\usepackage{dcolumn}
\usepackage{bm}
\usepackage{natbib}
\usepackage{upgreek}
\usepackage{amsmath}
\usepackage{amssymb}

\newcommand*{\cm}[1]{#1~cm$^{-1}$}

\begin{document}

\title{Terahertz properties of Dirac fermions in HgTe films with optical doping.}

\author{V. Dziom}
\author{A. Shuvaev}
\affiliation{Institute of Solid State Physics, Vienna University of
Technology, 1040 Vienna, Austria}
\author{N. N. Mikhailov}
\affiliation{ Novosibirsk State University, Novosibirsk 630090, Russia}
\author{A. Pimenov}
\affiliation{Institute of Solid State Physics, Vienna University of
Technology, 1040 Vienna, Austria}

\begin{abstract}
Terahertz properties of mercury telluride (HgTe) films with critical thickness are
presented and discussed. The density of the charge carriers is controlled using contact-free optical doping by visible light. In the magneto-optical response of HgTe
the contribution of two types of carriers (electrons and holes)
can be identified. The density
of the electrons can be modified by light illumination by more than one order of magnitude.
As the hole density is roughly illumination-independent, the terahertz
response of the illuminated samples becomes purely electronic. In some cases,
light illumination may switch the qualitative electrodynamic response from
hole-like to the electron-like. The cyclotron mass of the electrons could be
extracted from the data and shows a square root dependence upon the charge
concentration in the broad range of parameters. This can be interpreted as a clear proof of a linear dispersion relations, i.e. Dirac-type charge carriers.
\end{abstract}

\date{\today}

\maketitle

\section{Introduction}

Physical properties of relativistic Dirac states~\cite{hasan_rmp_2010, qi_prb_2008} have attracted much
interest recently, as they exhibit a number of unusual and nontrivial
electrodynamic properties.
These effects arise from linear dispersion of the charge carries known as 
Dirac cone. Within a Dirac cone the cyclotron effective mass of the charge
carriers strongly depends upon the position of the Fermi level (as controlled
by the charge density) and vanishes at the center of the cone. Unusual
electrodynamics at the interface between classical and quantum physics is
expected as, e.g., a universal Faraday effect or an anomalous Kerr
rotation~\cite{tse_prl_2010, tse_prb_2011, maciejko_prl_2010, %
tkachov_prb_2011}. 

Among various materials the system HgTe is outstanding as it provides a universal tool to investigate several complementary effects within the same composition. Bulk HgTe is a zero-gap semiconductor
due to the degeneracy of the two hole bands at the gamma point. The energy
gap is finite for thin strained films, which leads to a freezing out of the
bulk carriers at low temperatures. Therefore, for strained bulk samples the pure
two-dimensional (2D) character of electrons can be expected, which allows to
search for unusual electrodynamics as predicted by theory of topological insulators. In static transport
experiments, a strained HgTe layer exhibits a quantum Hall effect
(QHE)~\cite{brune_prl_2011}, yielding direct evidence that the charge
carriers in these layers are confined to the topological two-dimensional (2D)
surface states of the material.

The bulk HgTe is characterised by an inverted band structure. This means that
the $\Gamma_6$ band (which in conventional semiconductors is a conduction band) lays below the $\Gamma_8$ bands (which are normally the light- and heavy-holes bands). In such a case the $\Gamma_6$ band is a completely filled valence band, the heavy-holes subband of $\Gamma_8$ is a valence band and the light-holes subband of $\Gamma_8$ is the conduction band. As the light- and heavy-holes bands are degenerate at the center of the Brillouin zone, HgTe is a zero-gap semiconductor. As the sample thickness is decreased, the $\Gamma_6$ band rises in the energy and at the critical thickness of 6.3~nm passes over the $\Gamma_8$ bands. At smaller thicknesses, $\Gamma_6$ band is located above the $\Gamma_8$ bands and HgTe becomes a "conventional" semiconductor with the non-zero gap. Exactly at the critical thickness of 6.3~nm, when the $\Gamma_6$ and the light
holes from the $\Gamma_8$ bands are touching each other, they form a Dirac cone.
Therefore, the Dirac-like dispersion is a bulk band-structure effect in HgTe films with critical thickness~\cite{buttner_nphys_2011}. 

Magneto-optical experiments in semiconductor films provide a well-established tool to investigate the charge dynamics in external magnetic fields \cite{palik_rpp_1970}. Earlier this technique was successfully applied to investigate the complicated band structure in HgTe single crystals. More recently, the magneto-optics especially in the terahertz range have been utilized to study the two- and three-dimensional conducting states in graphene, Bi$_2$Se$_3$ and HgTe~\cite{shuvaev_prl_2011, hancock_prl_2011, shuvaev_prb_2013}. Compared to transport methods, optical measurements have the advantages of being contact-free and of directly accessing the effective mass via the cyclotron resonance. Both the presence of contacts and the patterning could lead to substantial changes of carrier concentration and of the position of the Fermi level. In the case of topological insulators these are very crucial parameters as the most interesting phenomena are expected in the vicinity of the Dirac point. The ability to observe the cyclotron resonance is another advantage of the magneto-optical technique. The cyclotron mass of the charge carriers $m_c$ can be determined from the resonance frequency $\omega_c$ via $m_c = eB/\omega_c$ and it is directly connected to the band structure near the Fermi level. Here $e$ is the electron charge and $B$ is the external magnetic field.

In experiments on the electrodynamics of topological insulators, the control of the charge density is very important. Such parameter is necessary to shift the Fermi level between electron- and hole-conduction within the Dirac cone. A classical tool to achieve this goal to use a  transparent gate electrode to change the charge concentration.  Recently, it has been demonstrated that a gated topological insulator may effectively be used to control the intensity and polarization of the terahertz radiation at room temperature~\cite{shuvaev_apl_2013}.

\begin{figure}[tbp]
\centerline{\includegraphics[width=0.7\columnwidth,clip]{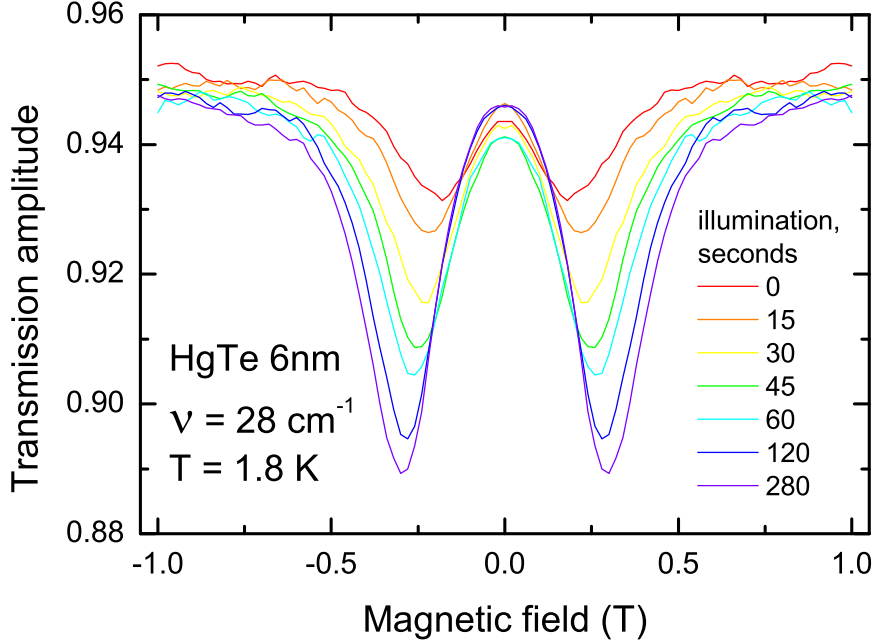}}
\caption{Demonstration of the charge control by light illumination in HgTe. The minima in transmission around $\sim 0.3$~T correspond to the cyclotron resonance of the Dirac-like carriers. The deepening of the minima under illumination reflects the increase of the charge density. The shift in position of the minima indicates the change in effective mass according to the expression $m_c = eB/\omega_c$.}
\label{ftran}
\end{figure}

Here we concentrate on an alternative route to modify the charge density in
topological insulators: light illumination (Figure~\ref{ftran}). This method
is well established in HgTe semiconductors with both parabolic and linear
dispersions~\cite{kvon_jetpl_2011, ikonnikov_sst_2011, olbrich_prb_2013, %
zoth_prb_2014} and is based on specific
properties of the dispersion relations in this material. At low temperatures
the channel of recombination of the photoinduced electrons is forbidden by
the momentum conservation. The light illumination has several advantages
compared to the gate method: the surface of the sample remains clear and
the charges are distributed more homogeneously along the surface and in the
bulk.

\subsection*{HgTe films}

The samples presented in this paper are epitaxial mercury telluride films with
the thickness of  6.3~nm grown on the [013] GaAs substrate. This thickness has been demonstrated to reveal a critical value for which the dispersion relations of electrons form a Dirac cone~\cite{bernevig_science_2006, konig_science_2007}.

An important feature of the HgTe samples is the ability to tune the charge carriers
density via induced photoconductivity. The visible light illumination causes
the electron density to increase. At low temperatures such additional charge carriers
are sustained for the long time period (of the orders of weeks) and the
temperature cycling up to room temperatures is needed in order to relax
these carriers back to their impurities centers. Experimentally, the dark
cooled sample has the one-way possibility to increase the charge concentration
via illumination. In our setup this was achieved by means of green light
LED mounted behind the nontransparent windows made of black paper (which is transparent for the terahertz radiation and opaque for infrared and visible light). The amount of additional
charge carriers brought into the conduction band is controlled by the
illumination time.

\section{Terahertz spectroscopy of cyclotron resonance}

Spectroscopic experiments in the terahertz frequency range (\cm{3}
$< \nu <$ \cm{30}) have been carried out in a Mach-Zehnder
interferometer arrangement~\cite{volkov_infrared_1985} which allows
measurements of the amplitude and the phase shift in a geometry with
controlled polarization of radiation. Theoretical transmittance
curves \cite{shuvaev_sst_2012} for various geometries were
calculated from the susceptibilities using Fresnel optical equations
for the complex transmission coefficient and within the Berreman
formalism \cite{berreman_josa_1972}.

The main results shown in this paper were obtained in the constant-frequency measurement
mode. In such case the frequency of the terahertz radiation is fixed
and both the transmission amplitude and the phase shift of the radiation
passing through the sample is measured as a function of the magnetic field.
With respect to the polarization of the radiation there are two main
geometries which were used in our work. (In both cases a wire-grid polarizer
was placed in front of the sample producing linearly polarized incidence wave.) In one case another wire-grid polarizer placed behind the sample was oriented parallel to the incident polarization. We denote this arrangement as parallel polarizers geometry. In the other case the second polarizer was oriented at $90^{\circ}$ with respect
to the first one. This layout is called crossed polarizers geometry.
Measuring the amplitude and the phase shift of the transmitted radiation in two geometries corresponds to full determination of the transmission matrix of the sample~\cite{shuvaev_sst_2012}.
It should be noted that for the case of trivial sample like isotropic
dielectric only signal in the parallel geometry is expected. Nonzero signal
in the crossed geometry is indicative of some sort of polarization rotation
or appearance of nonzero ellipticity after passing through the sample.

\begin{figure}[tbp]
\centerline{\includegraphics[width=0.9\columnwidth,clip]{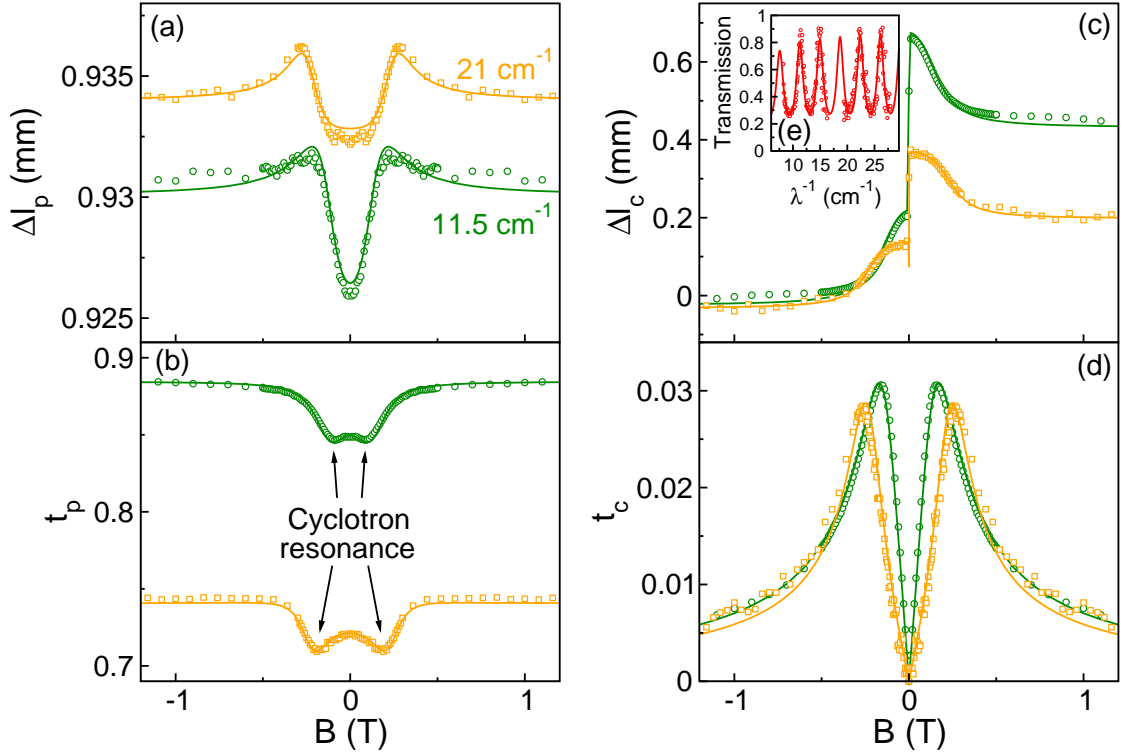}}
\caption{Magnetic field dependence of the transmission through 6.3~nm thin
HgTe sample \#1 after 45 seconds illumination time
in the parallel polarizers geometry (a), (b) and in the crossed
polarizers geometry (c), (d). The upper panels (a) and (c) show the phase
shift measured as the displacement of the movable mirror of the spectrometer,
the lower panels (b) and (d) show the transmission amplitude through the
sample. The inset (e) demonstrates the frequency dependence of the transmission in zero external magnetic field. Green open
circles are experimental data at \cm{11.5}, orange open squares - at \cm{21},
lines are fits using the Drude model as described in the text.}
\label{fscans}
\end{figure}

Typical measured data of HgTe films in parallel and crossed geometries are shown in Fig.~\ref{fscans}. The transmission amplitude in the
parallel geometry is shown in the lower left panel (b). Two distinct
symmetrical minima are clearly seen at low fields. They correspond to the
cyclotron resonance on free charge carriers in HgTe film. The minima at
\cm{21} are located at higher fields than the minima at \cm{11.5}, which
is in accordance with the linear dependence of cyclotron resonance upon
magnetic field: $\Omega_c \propto B$. For the case of charge carriers with
Dirac-like dispersion a nontrivial dependence of cyclotron resonance could
be expected: $\Omega_c \propto \sqrt{B}$. However this case is realized only
at high magnetic fields, when only few Landau levels are filled and the transitions between the Landau levels are observed separately.
In the present case several transitions between the Landau levels are overlapping, which leads to a recovery of the classical behaviour with $\Omega_c \propto B$. 

The lower right panel of Fig.~\ref{fscans} demonstrates the transmission amplitude in the crossed
polarizers geometry. The signal is zero without magnetic field, rises rapidly
in low fields reaching a maximum value and decreases upon further
increase of magnetic field. The emergence of the nonzero crossed signal is
the manifestation of the ac-Hall effect. Detailed analysis of the data including
the phase shift shows that both the rotation of the polarization and the
nonzero ellipticity of the radiation after passing through the sample are
present in HgTe. For a rather clean sample where the relaxation rate
of charge carriers is lower than the cyclotron frequency, $1/\tau < \Omega_c$,
the maximum in the crossed transmission signal also correspond to the position
of the cyclotron resonance.

Upper panels of Fig.~\ref{fscans} show magnetic field dependence of the phase
shift of the radiation after passing though the sample. The upper left panel
(a) corresponds to the parallel polarizers geometry and the upper right panel (c) to the
crossed polarizers
geometry. The phase shift is represented as a geometrical shift of the mirror
of the spectrometer needed to compensate for the phase shift caused by the
sample. This value corresponds to the change in the optical thickness of the sample. The phase shift in radians is then
$\Delta \varphi = 2 \pi \Delta l / \lambda$, where $\lambda$ is the wavelength
of the radiation. Whereas the experimental data in the parallel geometry in
the panel (a) show only relatively small changes, the data in the panel (c) for
the crossed geometry reveal an abrupt jump at zero magnetic field.
This is due to the fact that the crossed signal is changing sign
when the magnetic field sweeps from positive to negative values, which corresponds to the change in phase of exactly $\pi$, or the half
wavelength $\lambda / 2$ as in Fig.~\ref{fscans}(c).

The inset (e) in Fig.~\ref{fscans} demonstrates typical spectra in the
parallel polarizers geometry in zero magnetic field. The deep oscillations
are caused by a Fabry-P\'{e}rot-like multiple reflections within the transparent
substrate, which in the particular case was 0.387~mm thick undoped GaAs. The values of the maxima
in transmission are close to unity demonstrating high transparency of the HgTe
film. The decrease of their amplitude towards low frequencies is in accordance
with the Drude-like behaviour of the charge carriers, when their relaxation
rate is located in the experimental frequency range. In general, the
measurements with varying frequency are less accurate. This is caused by
numerous uncontrolled standing waves arising in the beam path due to reflections from the surfaces of optical components. Such standing waves give rise to non-reproducible changes in the amplitude of the signal upon the frequency sweep.
However this problem is absent when performing magnetic field sweeps at a fixed
frequency, which we have utilized in present experiments.

\subsection*{Drude model for magneto-optical transmission}

In order to analyze the magneto-optical data, a simple Drude model has been proved to provide an adequate description \cite{tse_prb_2011, tkachov_prb_2011, shuvaev_prb_2013}. Within
this model a sample of mercury telluride is modelled by an infinitely thin film
with a two-dimensional conductivity $\sigma_{2D} = \sigma_{3D} d_f$. Here,
$\sigma_{3D}$ is three-dimensional conductivity and $d_f$ is the film
thickness. In the case of nonzero magnetic field
normal to the film (along the $z$-axis) the conductivity is a $(2 \times 2)$ tensor with
all components different from zero:
\begin{eqnarray}
&& \sigma_{xx} = \sigma_{yy} = \frac{1 - \imath \omega \tau}
{(1 - \imath \omega \tau)^2 + (\Omega_c \tau)^2} \sigma_0 \,, \label{drude}\\
&& \sigma_{xy} = -\sigma_{yx} = \frac{\Omega_c \tau}
{(1 - \imath \omega \tau)^2 + (\Omega_c \tau)^2} \sigma_0 \,. \nonumber
\end{eqnarray}
Here, $\omega = 2 \pi \nu$ is the angular frequency of the terahertz radiation,
$\tau$ is the relaxation time of the charge carriers,
$\sigma_0 = n e^2 \tau / m_{eff}$ is the dc conductivity and
$\Omega_c = e B / m_{eff} $ is the cyclotron frequency.

A method to obtain the transmission spectra and the polarization state of the radiation in case of a thin conducting film on a substrate has been published previously \cite{shuvaev_prl_2011, shuvaev_sst_2012}. In case of transmission data only approximate final equations have been given and the matrix equations have been inverted numerically.
However, for a two dimensional conducting film on an isotropic dielectric
substrate the complex transmission can be obtained analytically. For a substrate with permittivity $\varepsilon$, the final equations for the spectra in parallel ($t_p$) and crossed ($t_c$) polarizers are given by:
\begin{eqnarray}
t_p &=& \frac{2 a_{xx}}{a_{xx}^2 + a_{xy}^2} \,, \label{transmission} \\
t_c &=& \frac{2 a_{xy}}{a_{xx}^2 + a_{xy}^2} \,, \nonumber
\end{eqnarray}
where
\begin{eqnarray}
a_{xx} &=& (1 + \sigma_{xx} d_f Z_0)(\cos(kd) -
    \imath Z \sin(kd)) + (\cos(kd) - \imath Z^{-1} \sin(kd)) \nonumber \\
a_{xy} &=& \sigma_{xy} d_f Z_0 (\cos(kd) - \imath Z \sin(kd)) \nonumber
\end{eqnarray}
Here $d_f$ is the conducting film thickness, $d$ is the thickness of
the substrate, $Z_0 \approx 377$~Ohm is the impedance of free space,
$Z = 1/\sqrt{\varepsilon}$ is the relative impedance of the substrate and
$k = \sqrt{\varepsilon} \omega / c$ is the wavevector of the radiation
in the substrate. As can be clearly seen, Eqs.~(\ref{transmission}) can be inverted analytically to get the solution for the complex conductivity. The final expressiond are a bit lengthy and they will not be presented here.

\section{Results and Discussion}

In the spectra shown in Figs.~\ref{ftran},\ref{fscans} only a single cyclotron resonance is observed. It is therefore reasonable to use only
one type of charge carriers in
the description. Besides, from the sign of the phase shifts it may be derived directly that the dominating carriers are negatively charged, i.e. they are electrons. The solid lines in Fig.~\ref{fscans} are model
calculations according to Eqs.~(\ref{drude},\ref{transmission}) superimposed
with the data for 45 seconds illumination time. Good quality of these simultaneous fits of four data sets supports the validity of the approximation. However, as will be seen below, in several cases the data suggests the presence of a second type of carriers which in few cases may even dominate the spectra.

\begin{figure}[tbp]
\centerline{\includegraphics[width=0.75\columnwidth,clip]
{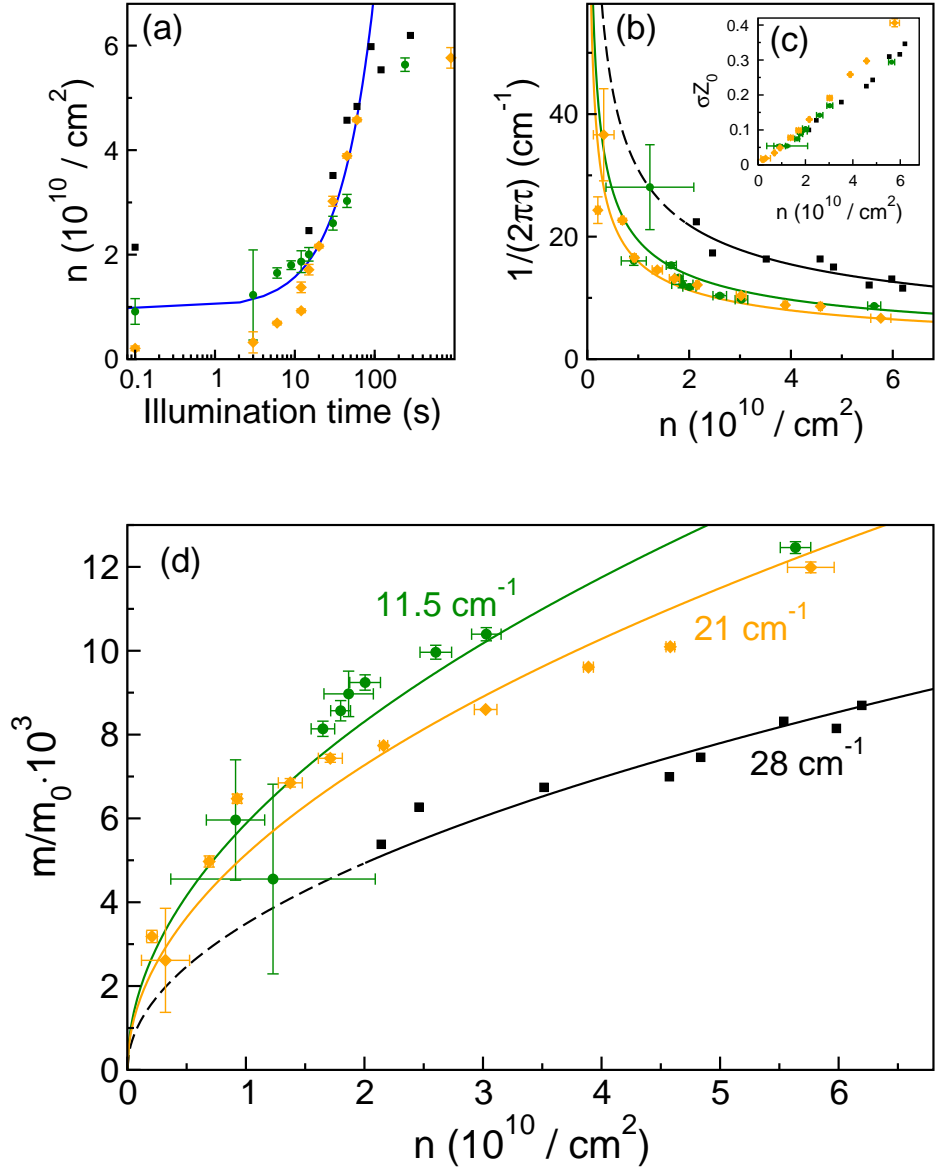}}
\caption{Fit parameters obtained from the complex transmission data in HgTe films.
Black squares are parameters of the sample \#2 measured at \cm{28}, green
circles denote sample \#1 measured at \cm{11.5} and orange diamonds are
for sample \#1 at \cm{21}.
Panel (a) shows the dependence of the charge carrier concentration (electrons)
upon the illumination time. The data for the dark sample with no illumination is
shown at 0.1~s in order to fit into the logarithmic scale. Solid line corresponds to a linear fit. Panel (b)
demonstrates the relaxation rate $1/2 \pi \tau$ as a function of the carriers
concentration $n$. Solid lines demonstrate the $1/2 \pi \tau \sim 1/\sqrt{n}$ behavior. The dimensionless 2D conductivity $\sigma Z_0$
vs. concentration is presented in the inset (c). Here, $Z_0 \approx 377$~Ohm is
impedance of vacuum. The lower panel (d) demonstrate the dependence
of the electron effective mass upon their concentration. The lines are
square root fits of the experimental parameters.}
\label{params_electrons}
\end{figure}

The parameters of the electrons in HgTe as obtained from the fits are shown in Fig.~\ref{params_electrons}. The upper left
panel (a) demonstrates the dependence of the charge carriers density $n$
against illumination time in the logarithmic scale. In order to accommodate
the dark sample with the illumination time 0, it is formally denoted by the time
of 0.1 seconds in the figure. The green circles are parameters for the sample \#1
at \cm{11.5}, the orange diamonds - for the same sample at \cm{21} and the
black squares are parameters of sample \#2 at \cm{28}. The change of the
concentration of at least one order of magnitude is clearly achievable in the present
experiment. As the illumination time is a parameter which is very specific for
the particular setup and generally has limited meaning, in the following the charge carriers
density $n$ will be used in the plots as a tuning parameter.

The charge carrier relaxation rate $1/2 \pi \tau$ is shown in the upper right
panel (b) of Fig.~\ref{params_electrons} as a function of charge carrier
density $n$. The  fit values for the same sample \#1 at two
different frequencies of \cm{11.5} (green circles) and \cm{21} (orange
diamonds) coincide rather well within the experimental accuracy (shown by
error bars). An increase of the relaxation rate toward low carrier densities
is well known in semiconductor physics and
is may be explained by the decreasing screening of the random potential by the charge
carriers. At low carrier densities the effective cyclotron mass becomes
very low and the cyclotron resonances are not resolved anymore. In this
case the determination of the parameters of the carriers becomes unstable
and results in large error bars. The black squares show the fit results
for the sample \#2 at \cm{28}. Here a systematic shift of relaxation rate
towards higher values compared to the sample \#1 can be attributed to the uncontrolled changes during the
sample preparation.

The inset (c) in the upper right panel of Fig.~\ref{params_electrons}
demonstrates almost linear dependence of the static conductivity
$\sigma_0$ on charge carriers density $n$. The linear character
of the curves and the fact that they closely coincide for both samples
indicate that the mobility $\mu$ of the charge carriers is constant
across the samples and the density ranges. This is evident from the
formula for the conductivity $\sigma = n e \mu$.

The concentration dependence of the effective electron mass in the HgTe films is shown in the panel (d) of Fig.~\ref{params_electrons}. For both samples this dependence follow the square root law $m_c \sim \sqrt{n}$. Especially for sample \#1 (green circles and orange diamonds) the square root behavior can be observed in extremely broad range of densities. The absolute values of the effective electron mass deviates only slightly for two frequencies, \cm{11.5} and \cm{21}. This may be an indication that that the magnetic field dependence of the cyclotron frequency starts to deviate from the linear low-field regime and an influence of the high-field $\Omega_c \propto \sqrt{B}$ regime is visible.
 
The square root behaviour is characteristic for the carriers with the Dirac-like dispersion relations $E=\hbar \upsilon_F |k|$ as observed in graphene~\cite{novoselov_nature_2005}. Here $\upsilon_F=const$ is the Fermi velocity.
From the expressions for the two-dimensional density $n_{2D}=k_F^2/(4 \pi)$ and quantum mechanical and classical definitions of the the cyclotron frequency~\cite{ashcroft_book,  hancock_prl_2011, ariel_arxiv_2012} $\omega_c = eB\upsilon_F/(\hbar k_F)$ and $\omega_c = eB/m_c$, respectively, one get $m_c=\hbar \sqrt{4 \pi n}/\upsilon_F$. These estimates clearly support the behaviour observed in Fig.~\ref{params_electrons}(d) and they correlate well with the results at higher frequencies~\cite{kvon_physe_2008, ikonnikov_jetpl_2010, ikonnikov_sst_2011, olbrich_prb_2013, zoth_prb_2014}.

\begin{figure}[tbp]
\centerline{\includegraphics[width=0.65\columnwidth,clip]{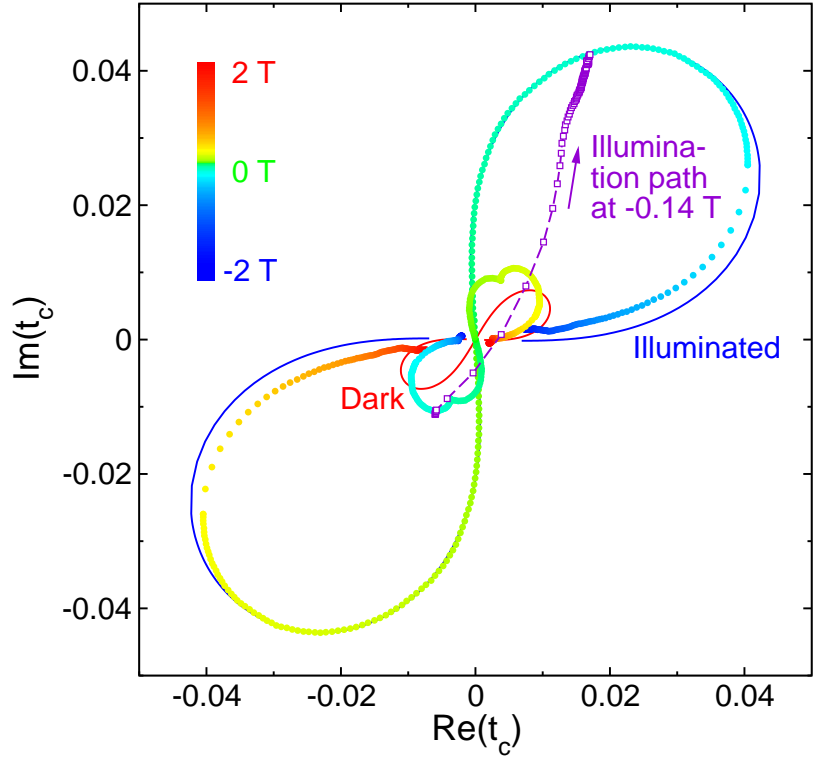}}
\caption{Polar plot of the transmission through the sample \#1 in the crossed
polarizers geometry as a function of magnetic field at \cm{11.5}.
The complex data is reconstructed from the amplitude
and the phase shift of the transmission (see Fig.~\protect\ref{fscans}). Solid circles represent the experimental data as indicated, open squares demonstrate the illumination
process taken at $B=-0.14$~T. The blue and red lines are fits within the Drude
model. Note the change of the orientation of the lobes between the dark and
the illuminated sample corresponding to the change from the hole-like to the electron-like
charge carriers.}
\label{polar_holes}
\end{figure}

\begin{figure}[]
\centerline{\includegraphics[width=0.7\columnwidth,clip]{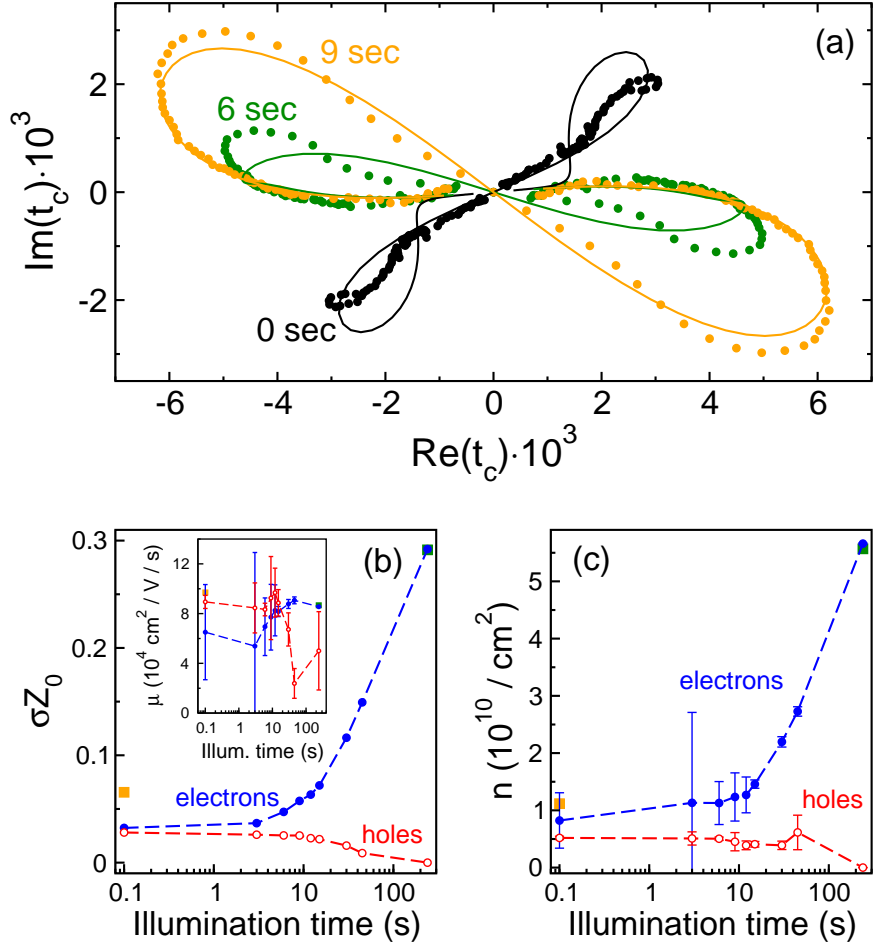}}
\caption{Panel (a) shows magnetic field dependence of transmission in crossed
polarizers geometry for sample \#1 at \cm{21}. The data are the same as in
Fig.~\protect\ref{fscans}, but represented in the polar plot form. Symbols
are experimental results, lines - fits within the Drude model using two types of
charge carriers. Strong deviations from the single carriers Drude model at
low illumination times are clearly seen. The dimensionless conductivity
$\sigma Z_0$ and the carrier concentrations $n$ as functions of illumination
time obtained from the fits are shown in panels (b) and (c), respectively.
Blue solid circles denote electrons, red open circles - holes. The green
and orange squares are fit parameters for electrons and holes as
obtained from the data in Fig.~\protect\ref{polar_holes}. The inset in (b) shows the mobilities of holes and electrons.}
\label{params_both}
\end{figure}

\subsection*{Hole contribution}

The results presented above were clearly dominated by a single electron contribution. Therefore, the description using only one type of carriers has led to reasonable interpretation of the data. As mentioned previously, in case where the electron contribution was small, additional details in the spectra could be detected. This can be interpreted as a contribution of a second type of carriers. 

In few experiments with samples in the dark even the dominating character of the hole contribution could be observed. An example of such spectra is presented in Fig.~\ref{polar_holes}. In this figure only the transmission in crossed polarizers is shown as it is mostly sensitive to fine details of the carrier contributions. The data are given in the complex plane plot such as $Im(t_c)$ is plotted as a function of $Re(t_c)$ including the phase information (which should be compared to conventional presentation in Fig.~\ref{fscans}). In such a plot the magneto-optical response of the charge carriers demonstrates a characteristic figure-eight shaped curve. Importantly, the sign of the charge carriers is directly obtained as the orientation of the curve in the complex plane.

As a typical example, purple open circles in Fig.~\ref{polar_holes} mark the position of $t_c$ on the complex plane at $B=-0.14$~T. It is clear, that the curve of the magneto-optical response in this case is inverted between the dark and the illuminated sample. Open squares in Fig.~\ref{polar_holes} show the path of $t_c( @  -0.14 \mathrm{T})$, i.e. the point which is close to one maximum of the response curve. In the dark case the sample shows clear orientation of the hole contribution. After switching on the light illumination the $t_c(-0.14 \mathrm{T})$ shifts to the origin (0,0) and then increases again with the opposite sign. This behaviour demonstrates the inversion from the hole to the electron contribution.

Solid lines in Fig.~\ref{polar_holes} demonstrate the results of the fitting to the measured data at fixed illumination time. The response on the electron side is well fitted by the simple model. On the hole side, only a qualitative fit may be obtained (red line). In addition, the experimental curve shows some fine structure. This is an indication of the fact that further corrections to the hole response may be needed, e.g. inclusion of further charge carriers. However, possible additional contributions are weak and their parameters cannot be reliably extracted from the fits. Therefore, in the present discussion only two contributions to the magneto-optical response will be considered: electrons giving a main response and holes as a smaller correction.

The HgTe samples investigated in this work showed distinct memory effects. This resulted in the fact that the "dark" state of the sample could not be exactly reproduced. Possible reason for these observation is the existence of charge traps with hysteresis effects. Normally, in the dark state as obtained after cooling from room temperature and without light illumination, the contributions of holes and electrons were comparable. In this sense, clear visual separation as exemplified in Fig.~\ref{polar_holes} could be obtained only in the begining of the experimental series. Other example of the spectra in the dark state is shown in Fig.~\ref{params_both}(a) by black symbols. Here more complicates picture compared to Fig.~\ref{polar_holes} is seen. Fortunately, the overall behaviour of the experimental data could be reasonably described taking only two sorts of the charge carriers into account. The results of such fits are shown in Fig.~\ref{params_both}(a) as solid lines demonstrating that even a complicated behaviour may be qualitatively understood as a mixture of hole and electron contributions. The attempts to include more charge carriers into consideration did not led to stable fits. As the response of the holes is weak, only few parameters like conductivity, density, and mobility could be determined unambiguously. The effective mass and the scattering rate of holes contribute to the mobility simultaneously as $\mu = e \tau/m_{eff} $ and they could not be separated by the fitting procedure.

The static conductivity and the density of electrons and holes in HgTe in the approximation of two types of charge carriers are shown in Fig.~\ref{params_both}(b,c). As expected, the parameters of the electrons in the two-carriers fits remain basically the same as in Fig.~\ref{params_electrons} (electrons only). Within the experimental accuracy the density of the holes remains independent of illumination indicating that mobile hole states are not affected by light.

The Dirac cone is symmetrical with respect to positive and negative directions in energy. If the Fermi-level is above the Dirac point, the charge carriers are electron-like. If the Fermi-level in below the Dirac point, the charge carriers will have the holes-like character. Fermi level for the samples in the dark state is most probably determined by the preparation conditions and partly by the temperature/doping history of the sample. Electron and hole contributions in the present study are observed from the same cyclotron resonance mode. Therefore, we conclude that both types of carriers correspond to Dirac dispersion.

\section{Conclusion}

Terahertz properties of the mercury telluride thin films with critical thickness are investigated. Using optical doping by visible light illumination, the charge carrier concentration could be modified by more than one order of magnitude. In some cases, using light as a parameter may switch the qualitative electrodynamic response from hole-like to the electron-like. Especially towards low electron density the cyclotron mass shows a square root dependence upon the charge concentration. This can be interpreted as a clear proof of a linear dispersion relations, i.e. Dirac type carriers.

\subsection*{Acknowledgements}

We acknowledge valuable discussion with G. Tkachov, E. Hankiewicz, and Z. D. Kvon. This
work was supported by Austrian Science Funds (I815-N16, W-1243, P27098-N27).

\bibliography{literature}

\end{document}